\documentstyle[preprint,aps]{revtex}

\draft

\begin{document}

\title{Enhanced Transmission Due to Disorder}

\author{V.\ D.\ Freilikher, B.\ A.\ Liansky, I.\ V.\ Yurkevich}
\address{
The Jack and Pearl Resnick Institute of Advanced Technology,
Department of Physics,\\
Bar-Ilan University,
Ramat-Gan 52900, Israel
}

\author{A.\ A.\ Maradudin}
\address{
Department of Physics,
University of California,
Irvine, CA 92717, USA
}

\author{A.\ R.\ McGurn}
\address{
Department of Physics,
Western Michigan University,
Kalamazoo, MI 49008, USA
}

\date{\today}
\maketitle

\begin{abstract}
The transmissivity of a one-dimensional random system that is periodic on
average is studied. It is shown that the transmission coefficient for
frequencies corresponding to a gap in the band structure of the average
periodic system increases with increasing disorder while the disorder is
weak enough. This property is shown to be universal, independent of the type
of fluctuations causing the randomness. In the case of strong disorder the
transmission coefficient for frequencies in allowed bands is found to be a
non monotonic function of the strength of the disorder. An explanation for
the latter behavior is provided.
\end{abstract}


\narrowtext

It is well known today \cite{ref1} that in the propagation of a classical
wave through a one-dimensional periodic structure of finite length the
amplitude of the transmitted wave decreases exponentially with increasing
length of the system when the frequency of the wave is in a gap in the
photonic band structure of the infinite lattice of the same period. If this
periodic structure is now randomly disordered in such a way that it remains
periodic on average, it might seem as if a wave whose frequency lies in the
gap would be even more strongly attenuated, due to the additional random
scattering. In this paper we show, on the basis of three different models,
that in fact the opposite occurs: the transmissivity of the disordered
system is larger than that of the periodic structure, when the frequency of
a wave propagating through it falls in a band gap of the latter. A possible
reason for this counterintuitive result will be presented.

We consider a scalar wave $U(z)$ that satisfies the one-dimensional
Helmholtz equation
\begin{equation}
\label{eq1}\left[ \frac{d^2}{dz^2}+\frac{\omega^2}{c^2}\epsilon(z)%
\right]U(z)=0 ,
\end{equation}
where $\epsilon(z)$ is a random, periodic on average, function inside the
disordered region $z\in(0,L)$, and is equal to $\epsilon_0$ for $z\not%
\in(0,L)$. In the case when a plane wave of a frequency $\omega$ is incident
on the segment $(0,L)$ from the left the solution of Eq.\ (\ref{eq1}) for $z%
\not\in(0,L)$ can be written in the form
\begin{equation}
\label{eq2}U(z)=
\cases{
e^{ik_0z}+r(L)e^{-ik_0z}\hphantom{hlis}\text{ for }z<0\cr
t(L)e^{ik_0z}\hphantom{thisthisthis\,\,}\text{ for }z>L ,\cr
}
\end{equation}
where $k_0=\epsilon^{1/2}_0\omega /c$, which defines the reflection
and transmission coefficients $r(L)$ and $t(L)$, respectively.

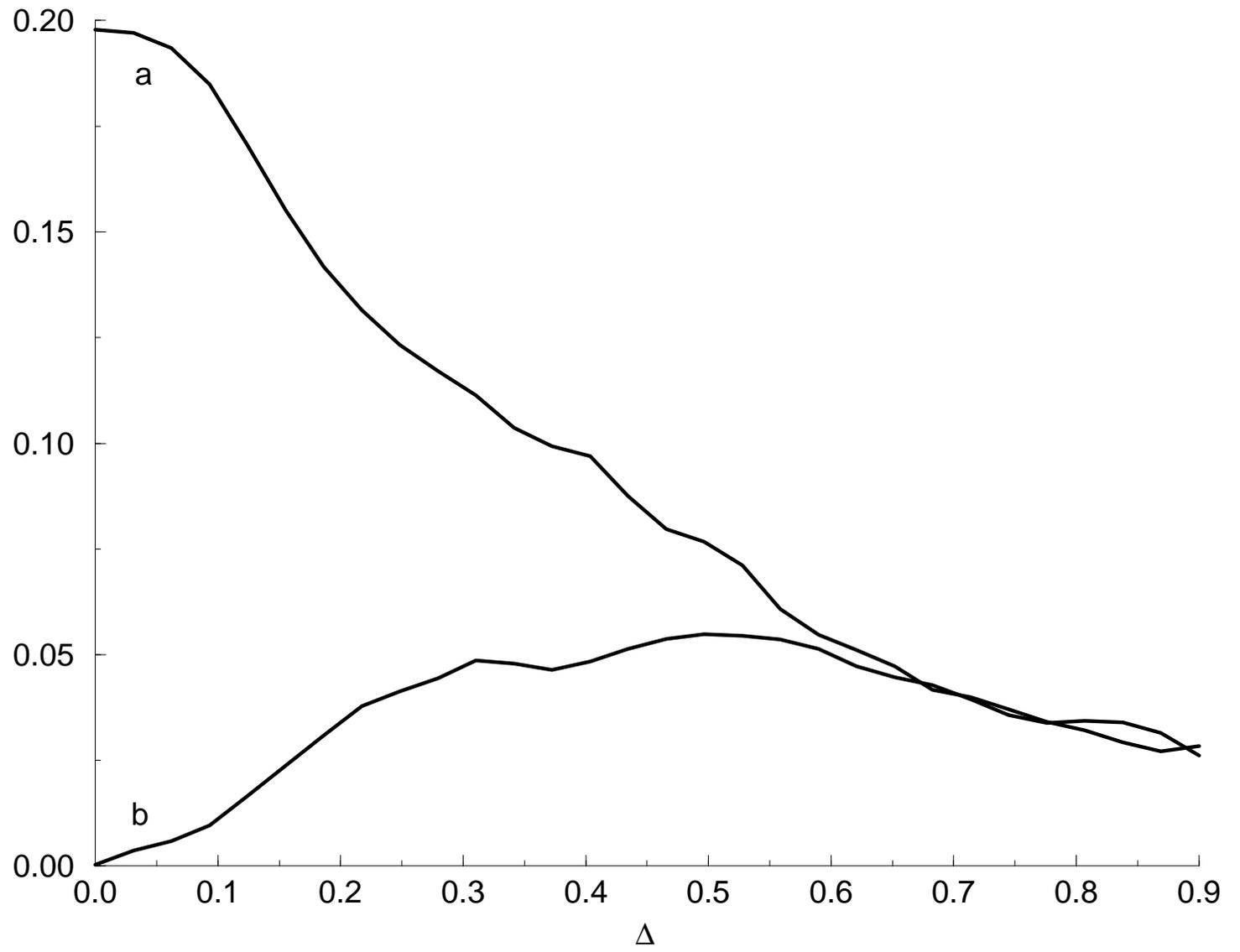
\begin{figure}
\caption{
Liapunov exponent as a function of the disorder parameter $\Delta$
for the first model with ``positional'' disorder. Curve {\it a}
corresponds to the wave frequency $\omega a/c=0.90$ above the lower
frequency edge of the first band gap $(\omega_-a/c = 0.873)$, and {\it b}
 corresponds
to the frequency $\omega a/c=0.85$ below the edge.
}
\label{layered1}
\end{figure}

The first model of a disordered structure that we consider is an alternating
array of $2N$ dielectric slabs of dielectric constants $\epsilon_1$ and $%
\epsilon_2$ in the region $(0, L)$. The width of the $i$th slab is given by $%
a_i=a(1+\Delta_i)$, where the $\Delta_i$ are independent random variables
that are uniformly distributed in the interval $(-\Delta, \Delta)$, where
clearly $0\leq\Delta < 1$. This structure is imbedded in a homogeneous
medium, and a wave of frequency $\omega$ is incident normally on it from the
left. The complex transmission coefficient, $t$, has been calculated as a
function of the length $L$ of the disordered structure by means of a
transfer matrix approach \cite{ref2}, for two frequencies of the incident
wave in the vicinity of the lower edge of the lowest frequency band gap in
the band structure of the average periodic lattice. For the parameters
assumed ($\epsilon_1=2, \epsilon_2=3.5$) the frequencies of the lower and
higher frequency edges of this band gap are defined by $\omega_-a/c=0.873$
and $\omega_+a/c=1.038$, respectively. In Fig.\ \ref{layered1} we plot the
large $L$ limit of $-(a/L)\langle \ln T\rangle=a/l(\omega)$ as a function of
the disorder parameter $\Delta$ for two different frequencies $\omega$ of
the incident wave. Here $T=\left|t\right|^2$ is the transmissivity of the
structure, the angle brackets denote an average over 100 realizations of the
structure, and $l^{-1}(\omega)$ is the Liapunov exponent.

{}From the results presented in Fig.\ \ref{layered1} we see that $%
l^{-1}(\omega)$ decreases monotonically with increasing $\Delta$ for the
frequency of the electromagnetic wave within the band gap, which in turn
means that the transmittance of the structure increases with increasing
disorder. It is as if in the presence of the disorder channels for the
propagation of waves in this frequency range that are closed in the absence
of the disorder open up due to the partial filling of the density of
photonic states in the gap region by the tails of this density of states
from the higher and lower frequency bands bordering the gap. However, for
the frequency in the allowed band, we obtain an at first sight surprising
nonmonotonic dependence of the Liapunov exponent on the strength of the
disorder $\Delta$. The initial increase of $l^{-1}(\omega)$ with increasing $%
\Delta$ is readily understood as the suppression of the transmission in this
frequency range due to the multiple scattering of the incident wave caused
by the disorder. The decrease in the value of $l^{-1}(\omega)$ for strong
disorder can be explained by a rather simple argument that will be presented
below. The merging of the curves $l^{-1}(\omega)$ for all values of $\omega$
for values of $\Delta$ larger than about $0.6$ is a reflection of the fact
that for such large values of $\Delta$ the scattering structure has lost all
traces of its underlying periodicity, as a consequence of which the band
structure has disappeared, and waves of all frequencies see the same
disordered medium that is now homogeneous on average.

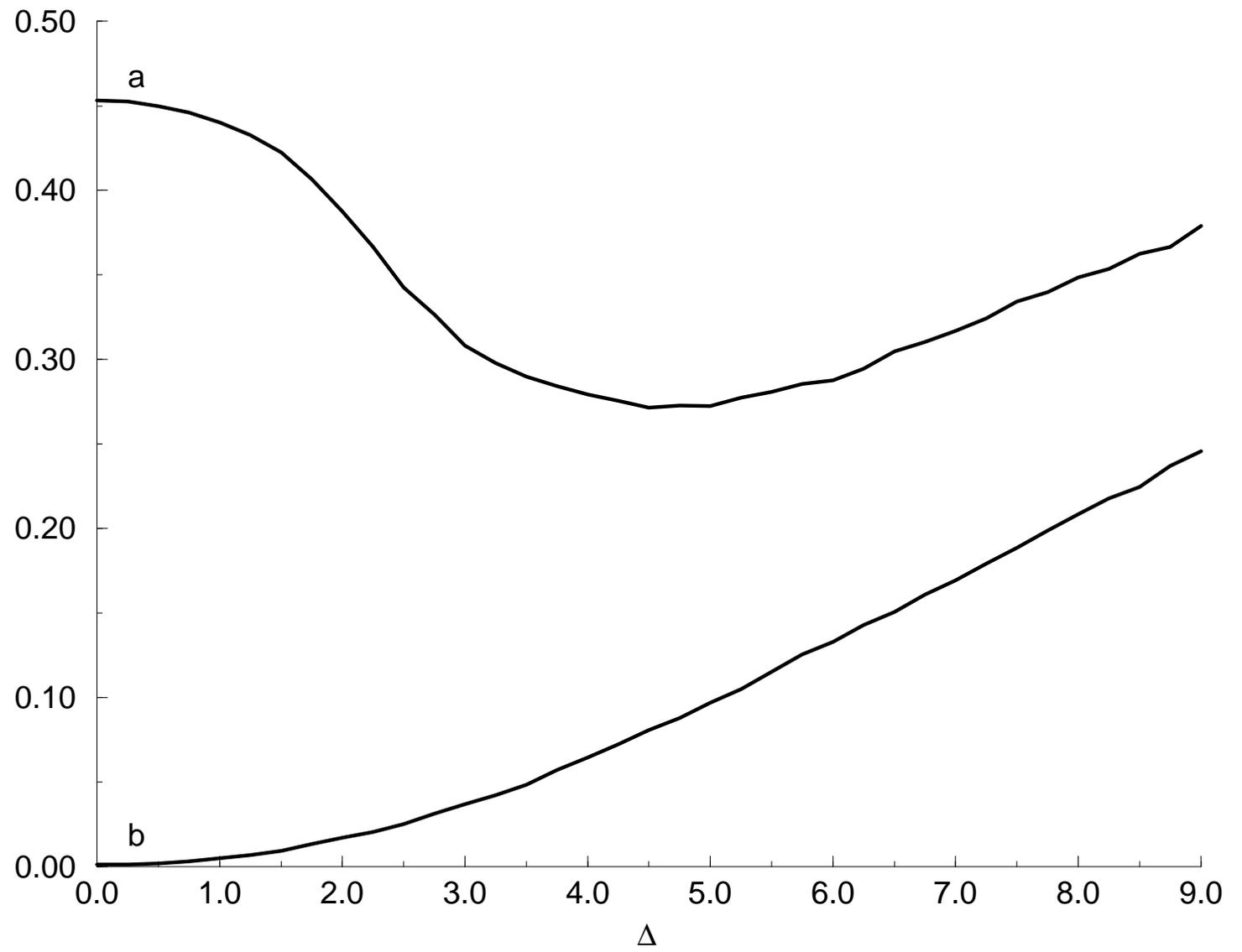
\begin{figure}
\caption{
Liapunov exponent as a function of the disorder parameter
$\Delta$ for the second model with random $\epsilon_i$. Curve {\it a}
corresponds to the wave frequency $\omega a/c=0.6$ above the lower
frequency edge of the first band gap, and {\it b} corresponds to
 the frequency $\omega a/c=0.5$ below the edge. For the  parameters
of the second model  assumed ($\epsilon=9.0$) the position of  the lower
frequency
edge of the first band gap is at $\omega_-a/c=0.568$.
}
\label{layered2}
\end{figure}

The second model of a disordered structure that we consider is a system of $%
2N-1$ dielectric slabs, each of thickness $a$, in which the dielectric
constant $\epsilon_i$ of $i$th slab is given by $\epsilon_{2i-1}=\epsilon+
\delta\epsilon_i, \epsilon_{2i} = 1$, where the $\delta\epsilon_i$ are
independent random variables that are uniformly distributed in the interval $%
[-\Delta, \Delta]$. In contrast with the first model of a disordered
structure considered, in which the disorder parameter $\Delta$ was limited
by physical considerations to the values $0\leq\Delta<1$, there is no
restriction on the (positive) values that $\Delta$ can take in the present
model. In Fig.\ \ref{layered2} we plot the dependence of the Liapunov
exponent $l^{-1}(\omega)$ as a function of the disorder parameter $\Delta$
when the frequency of the incident wave is in the lowest frequency gap of
the photonic band gap structure of the average periodic system, and when
it is in the allowed band below it. If the frequency is in the band gap, we
see that when $\Delta$ is increased from zero, $l^{-1}(\omega)$ initially
decreases, indicating that the transmittance $T=|t|^2$ is increasing with
increasing disorder. However, when $\Delta$ reaches a value of approximately
$4.5$, $l^{-1}(\omega)$ begins to increase monotonically with a further
increase of $\Delta$. In contrast, when the frequency of the incident wave
is in the allowed band $l^{-1}(\omega)$ increases monotonically with
increasing $\Delta$.

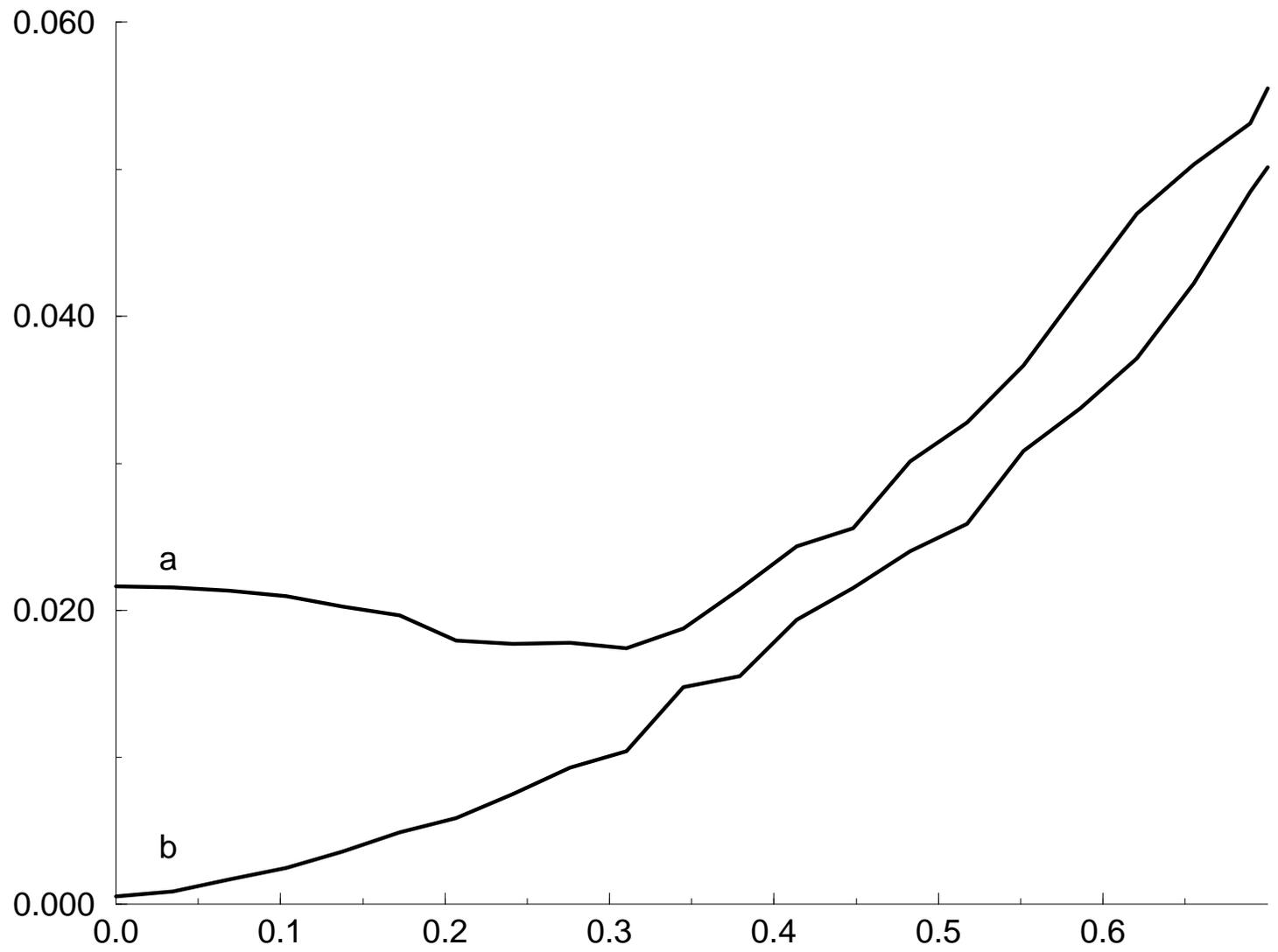
\begin{figure}
\caption{
Liapunov exponent as a function of the root mean square
of the random function $\delta\epsilon(z)$ in the third, continuous,  model.
The curves {\it a}
and {\it b} correspond to  wave frequencies above and below
 the lower frequency edge of the first band gap, respectively.
}
\label{imbedding}
\end{figure}

The third model of a periodic system with disorder considered in the
present paper is a continuous one: its dielectric constant is given by
\begin{equation}
\epsilon (z)=A\cos qz+\delta \epsilon (z),
\end{equation}
where $\delta \epsilon (z)$ is a zero-mean, Gaussian random process, with
the correlation function
\begin{equation}
\left\langle \delta \epsilon (z)\delta \epsilon (z^{\prime })\right\rangle
=\delta ^2\exp [-|z-z^{\prime }|^2/l_{\text{cor}}^2].
\end{equation}
For a numerical calculation of the transmission coefficient we used the
exact invariant imbedding equations \cite{refimbed1,refimbed2}
\begin{mathletters}
\label{eq:imbedd}
\begin{equation}
\frac{dr(L)}{dL}=
\frac{i}{2}k_0\epsilon(L)\left[e^{-ik_0L}+r(L)e^{ik_0L}\right]^2~
\end{equation}
\begin{equation}
\frac{dt(L)}{dL}=
\frac{i}{2}k_0\epsilon(L)t(L)\left[1+r(L)e^{2ik_0L}\right]~,
\end{equation}
\end{mathletters}
subject to the initial conditions $r(0)=0$, $t(0)=1$. The results are
presented in Fig.\ \ref{imbedding}.

When Figs.\ \ref{layered1}, \ref{layered2}, and \ref{imbedding}
are compared, it is apparent that if the disorder is weak enough, the
transmissivity for frequencies corresponding to gaps exhibits a rather
unusual universal feature independent of the type of fluctuations: the
transmission coefficient $T(\omega)\sim e^{-L/l(\omega)}$ increases when
disorder appears ($\Delta\not=0$), and continues to increase when the
fluctuations grow. For strong disorder ($\Delta\sim 1$ in the first model, $%
\Delta>4.5$ in the second one) the behavior of $T$ as a function of $\Delta$
depends drastically on the type of disorder. Along with the expected
decreasing of $T$ in the second and third models one can see a quite
surprising increase of the transmissivity of the system with ``positional''
discrete disorder (first model) when the strength of disorder $\Delta$
increases.

To understand the differences of the dependence of $l^{-1}(\omega )$ on the
parameter $\Delta $ in the two discrete models studied in the present work,
let us consider the disordered segment $(0,L)$ as a set of random
scatterers, each of which is a dielectric slab with a random width (first
model) or a random dielectric constant $\epsilon _i$ (second model) that
possesses random reflection and transmission amplitudes $R_{\text{ind}}$ and
$T_{\text{ind}}$, respectively. In addition to the randomness in the
individual scattering characteristics, we have in the first model random
distances $\delta _i=$$a(1+\Delta _{2i})$ between consecutive scatterers.
Then the transfer matrix $\tensor{M}$ for the entire system can be written
as the product $\tensor{M}=\tensor{M}_1\tensor{F}_1\tensor{M}_2\tensor{F}%
_2\cdots \tensor{M}_{N-1}\tensor{F}_{N-1}\tensor{M}_N$, where
\begin{equation}
\tensor{M}_i=\left(
\matrix{
\alpha_i & \beta_i \cr
\beta^*_i & \alpha^*_i \cr
}\right) \text{ and }\tensor{F}_i=\left(
\matrix{
e^{ik\delta_i} & 0 \cr
0 & e^{-ik\delta_i} \cr
}\right)
\end{equation}
are the transfer matrices for the $i$th scattering slab and for the
homogeneous space between consecutive scattering slabs, respectively. The
effect of the interfaces in this system are incorporated into the definition
of the matrices $\left\{ \tensor{M}_i\right\} $. The $11$-element of the
product $\tensor{M}\tensor{M}^{+}$ is related to the transmissivity of the
structure $T=|t|^2$ by
\begin{equation}
\label{mm11}\left( \tensor{M}\tensor{M}^{+}\right) _{11}=\frac 1T-1~.
\end{equation}
In averaging the matrix $\tensor{M}\tensor{M}^{+}$ we can use the
independence of the matrices with different subscripts. This means that in
the first model we can average over the distance between the $(N-1)$th and $N
$th scatterers independently of all the other scatterers:
%\onecolumn
\widetext
\begin{eqnarray}
\left\langle\tensor{F}_{N-1}\tensor{M}_{N-1}
\tensor{M}^+_N\tensor{F}^+_{N-1}\right\rangle
&&=
\left(\matrix{
\left\langle
\left|\alpha_N\right|^2+\left|\beta_N\right|^2
\right\rangle &
\left\langle
2\alpha_N\beta_N
e^{i2k\delta_{N-1}} \right\rangle \cr
\left\langle
2\alpha^*_N\beta^*_N
e^{-i2k\delta_{N-1}} \right\rangle  &
\left\langle
\left|\alpha_N\right|^2+\left|\beta_N\right|^2
\right\rangle \cr
}\right)\nonumber\\
&&\nonumber\\
&&=\left(\matrix{
\left\langle\displaystyle\frac{1+R_N}{T_N}\right\rangle &
2\left\langle\alpha_N\beta_N\right\rangle
\left\langle e^{i2k\delta_{N-1}}\right\rangle \cr
2\left\langle\alpha_N\beta_N\right\rangle^*
\left\langle e^{-i2k\delta_{N-1}}\right\rangle &
\left\langle\displaystyle\frac{1+R_N}{T_N}\right\rangle \cr
}\right) .
\label{scatterer}
\end{eqnarray}
%\nobreak
%\twocolumn
\narrowtext
If the disorder is strong enough ($2\Delta \cdot ka\gg 1$), we can neglect
the off-diagonal terms on the right hand side of Eq.\ (\ref{scatterer}). By
repeating this procedure for each of the random transfer matrices, we obtain
\begin{equation}
\label{mm11a}\left\langle \tensor{M}\tensor{M}^{+}\right\rangle
=\left\langle \frac{1+R_{\text{ind}}}{T_{\text{ind}}}\right\rangle ^N\left(
\matrix{
1 & 0 \cr
0 & 1 \cr
}\right) ~.
\end{equation}
{}From a comparison of Eqs.\ (\ref{mm11}) and (\ref{mm11a}) we find that
\begin{equation}
\left\langle \frac 1T\right\rangle =\frac 12\left[ 1+\left\langle \frac{1+R_{
\text{ind}}}{T_{\text{ind}}}\right\rangle ^N\right] ~.
\end{equation}
For a long system ($N\rightarrow \infty $) Eq. (10) becomes approximately
\begin{equation}
\left\langle \frac 1T\right\rangle \cong \frac 12\exp \left\{ N\ln
\left\langle \frac{1+R_{\text{ind}}}{1-R_{\text{ind}}}\right\rangle \right\}
{}~,
\end{equation}
which means that the localization length $l(\omega )$ is of the order of
\begin{equation}
\frac a{l(\omega )}=-\frac 1N\left\langle \ln T\right\rangle \cong \ln
\left\langle \frac{1+R_{\text{ind}}}{1-R_{\text{ind}}}\right\rangle ~.
\end{equation}
In obtaining this result we have approximated $\langle \ln T\rangle $ by $%
-\ln \langle 1/T\rangle $, and have ignored questions of the self-averaging
or non-self-averaging of the latter quantity to obtain a simple qualitative
result. For weak individual scattering we obtain
\begin{equation}
\label{raver}\frac a{l(\omega )}\approx 2\left\langle R_{\text{ind}%
}\right\rangle ~.
\end{equation}
Therefore in the case of strong disorder the transmission coefficient of a
disordered random system is determined completely by the mean value of the
reflection coefficient of a single scatterer. In the first model $R_{\text{%
ind}}$ is a periodic function of the width $d$ of the single slab, and the
averaging in Eq.\ (\ref{raver}) means just the integration of $R_{\text{ind}}
$ over $d$ in the interval $\left( a(1-\Delta ),a(1+\Delta )\right) $. It is
clear that for $\Delta \sim 1$ (strong ``positional'' disorder) the increase
of the interval of integration of a periodic function causes the decrease of
$\left\langle R_{\text{ind}}\right\rangle $.

In the second model (random $\epsilon_i$) the increase of $\Delta$ enhances
the strength of a single scatterer and obviously leads to an increase of $%
\left\langle R_{\text{ind}}\right\rangle$.

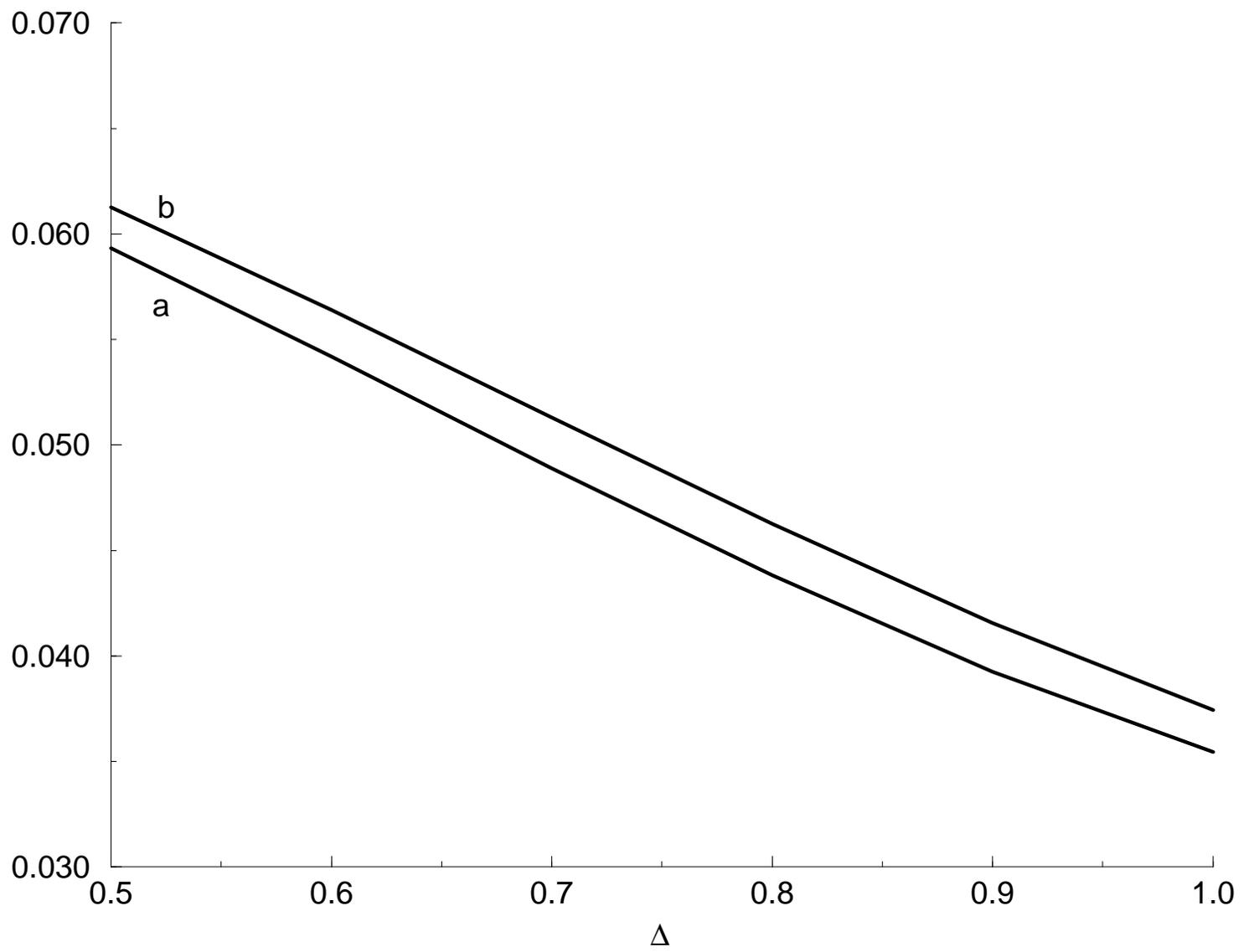
\begin{figure}
\caption{
Plot of $\left\langle R_{\text{ind}} \right\rangle$.
Curves {\it a} and {\it b} were calculated for the frequencies
that correspond to the
frequencies  of the curves {\it a} and {\it b} in Fig.\ 1.
}
\label{raverage}
\end{figure}

We have carried out independent numerical calculations of $\left\langle R_{
\text{ind}}\right\rangle$ (Fig.\ \ref{raverage}). The results are consistent
with the reasoning above, and provide an explanation for the behavior of $%
l^{-1}(\omega)$ for strong disorder (large $\Delta$) depicted in Fig.~\ref
{layered1}.

In conclusion, we have shown on the basis of three different models of a
one-dimensional random structure that is periodic on average, that the
transmissivity of the structure for waves with frequencies corresponding to
a gap in the band structure it possesses in the absence of the randomness is
increased by the randomness. In addition, we have given a qualitative
explanation of the observed nonmonotonic dependence of the length on the
disorder parameter $\Delta$ for waves whose frequencies are in an allowed
band.

\acknowledgments
This research was supported in part by the United States-Israel Binational
Science Foundation Grant No. 92-00248. The work of A.\ R.\ M.\
was supported in part by National Science Foundation Grant No.\
DMR 92-13793.

\end{document}